\documentclass[ArXiv,AXISWP,accept,moreauthors,pdftex,10pt,letterpaper]{axis} 

\usepackage{amsmath,amssymb}
\newcommand{\RomanNumeralCaps}[1]
    {\MakeUppercase{\romannumeral #1}}

\firstpage{1} 
\pubvolume{xx}
\issuenum{1}
\articlenumber{5}
\pubyear{2023}
\copyrightyear{2023}

\pdfoutput=1

\Title{Uncovering the First AGN Jets with AXIS}

\Author{Thomas Connor\orcidA{} $^{1,\dagger}$, Eduardo Ba\~nados\orcidB{} $^{2}$, Nico Cappelluti\orcidC{} $^{3}$, and Adi Foord\orcidD{} $^{4}$}

\AuthorNames{Thomas Connor, Eduardo Ba\~nados, Nico Cappelluti, Adi Foord}

\address{%
$^{1}$ \quad Center for Astrophysics $\vert$\ Harvard\ \&\ Smithsonian, 60 Garden St., Cambridge, MA 02138, USA\\
$^{2}$ \quad Max Planck Institute for Astronomy, K\"onigstuhl 17, 69117 Heidelberg, Germany\\
$^{3}$ \quad Department of Physics, University of Miami, Coral Gables, FL 33124, USA\\
$^{4}$ \quad Department of Physics, University of Maryland Baltimore County, 1000 Hilltop Cir, Baltimore, MD 21250, USA}

\firstnote{thomas.connor@cfa.harvard.edu} 

\abstract{Jets powered by AGN in the early Universe ($z \gtrsim 6$) have the potential to not only define the trajectories of the first-forming massive galaxies but to enable the accelerated growth of their associated SMBHs. Under typical assumptions, jets could even rectify observed quasars with light seed formation scenarios; however, not only are constraints on the parameters of the first jets lacking, observations of these objects are scarce. Owing to the significant energy density of the CMB at these epochs capable of quenching radio emission, observations will require powerful, high angular resolution X-ray imaging to map and characterize these jets. As such, \textit{AXIS} will be necessary to understand early SMBH growth and feedback.
 \emph{This White Paper is part of a series commissioned for the AXIS Probe Concept Mission; additional AXIS White Papers can be found at the  \href{http://axis.astro.umd.edu/}{AXIS website} with a mission overview \href{https://arxiv.org/abs/2311.00780}{here}}}

\begin{document}
\tableofcontents
\listoffigures


\section{Introduction}

After decades of steady but limited progress, the lights of the dark ages of the early Universe are finally being seen. The brightest and most accessible of these objects are Active Galactic Nuclei (AGN), the extremely-luminous inner accretion regions surrounding the first-formed supermassive black holes (SMBHs). Early results from \textit{JWST} have already started a torrent of new discoveries \citep{2023ApJ...953L..29L, 2023ApJ...951L...4W, 2023ApJ...951L...5Y}, and the promises of future missions---of \textit{Euclid} and \textit{Roman}, of the 30-meter telescopes, of the ngVLA and the SKA---position the 2030's to be a decade in which we can answer some of the fundamental questions of the first black holes, chief among which is ``how were such massive objects able to form so quickly?''

Hand in glove with the evolution of these systems is the role of jets, beams of relativistic particles powered by the black hole and its accretion region. The energy liberated by jets can be extreme, dominating over radiative emission and physically shaping not just the AGN host galaxies themselves \citep[e.g.,][]{2020NewAR..8801539H} but even the environments in which they reside, to the scale of galaxy clusters. Even small effects in the beginning of the Cosmos can lead to significant changes 13 billion years later, and so no picture of galaxy formation and evolution or of SMBH growth and seeding can be complete without the details of how jets are at work.

In the following, we discuss how early jets can significantly impact the growth of early SMBHs and the observational challenges that have kept high-redshift jets from being well-studied. High angular resolution X-ray imaging with a low-background instrument will be necessary to address these issues, motivating the need for \textit{AXIS} \citep{2023_AXIS_Overview}.

\section{Jet-Assisted Black Hole Growth}
The problem of growing the first SMBHs is simple. We have observed hundreds of AGN at $z>6$, with measured black hole masses of $10^9$ to $10^{10}\ {\rm M}_{\odot}$ \citep{2016ApJS..227...11B, 2023ApJS..265...29B, 2019ApJ...884...30W, 2023ApJS..269...27Y}; however, even if we assume that they have been growing at the Eddington limit since a scant 100 Myr after the Big Bang, these black holes would have had to form from seeds with masses of order ${\sim}10^5$--$10^6\ {\rm M}_\odot$ \citep{2021ApJ...907L...1W}. Producing seed black holes with such mass is a significant challenge for theoretical and computational models. Conversely, lower-mass seeds are more easily generated (remnants of massive stars are the only well-established black hole formation mechanism), but they would require sustained super-Eddington growth to produce observed quasars, itself a separate significant challenge. 
\begin{figure}[!t]
\centering
\includegraphics[]{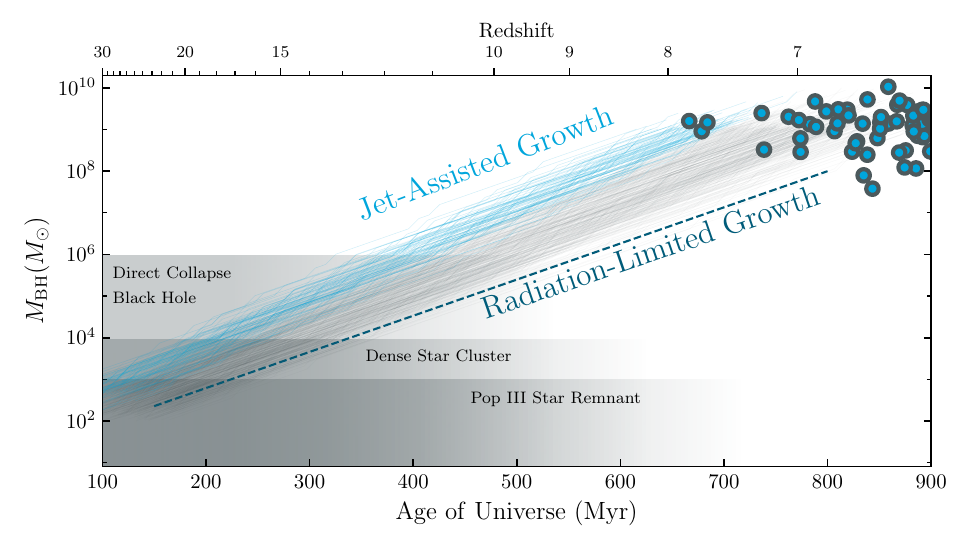}
\caption[Effect of Jets on Early Black Hole Mass Growth]{Mass and redshift distribution of the known high-redshift quasars with robust mass measurements (blue circles, \citep{2021ApJ...922L..24C}). The approximate mass distributions allowed by the three main seeding models are shown with horizontal gray bands. Simulating growth tracks from massive Pop \RomanNumeralCaps{3} star remnants with random periods of jet-enhanced accretion, a fraction of the light seeds can still produce the most extreme quasars observed (blue lines). In contrast, the radiation-only track (dashed line, Equation \ref{eqn:simple_eddington}) is not compatible with both stellar remnant seeds and the highest-redshift observed quasars. }\label{fig:mass_growth}
\end{figure}   

At the core of this conundrum is the Eddington limit. For a black hole to grow, it requires more material to be funneled through its accretion disk. Yet, further infall will increase the luminosity of the disk, thereby increasing the outward radiation pressure pushing against that same infall. This is characterized through the equation
\begin{equation} \label{eqn:simple_eddington}
\ln\left[({\rm M}(t)/{\rm M}(t_0)\right] \propto (t - t_0) \times (1 - \epsilon)/\epsilon,
\end{equation}
where $\epsilon$ is the accretion efficiency---that is, the fraction of infalling mass energy that does not end up inside the black hole. $\epsilon$ manifests as two separate hits to mass growth, owing to the loss of infalling mass (the $1-\epsilon$ term) and the energy released to resist infall (the $\epsilon$ of the denominator). For a radiative efficiency of $\epsilon=0.1$, Eddington-limited growth will require approximately 115 Myr to increase a decade in mass.

As to what could potentially form the observed SMBHs, there are, in general, three dominant theoretical models. The first is the simplest, in which a massive Population III star ends its life and leaves a massive remnant. Such seeds could, at their most optimistic extent, reach masses of order ${\sim}10^3\ {\rm M}_\odot$ \citep{2001ApJ...550..372F, 2014ApJ...781...60H, 2012ApJ...756L..19W}. For more massive remnants, one possible mechanism would involve mergers in a dense star cluster, whereby the hypermassive star could leave a remnant with a mass up to ${\sim}10^4\ {\rm M}_\odot$ \citep{2009ApJ...694..302D, 2017MNRAS.472.1677S}.  The third mechanism would be the collapse of atomic Hydrogen---potentially through a brief superstellar phase---directly into a black hole \citep{2013MNRAS.433.1607L, 2016MNRAS.463..529H, 2019Natur.566...85W}. For this direct collapse mechanism to work, however, it would require a plentiful source of Lyman--Werner photons to dissociate any latent ${\rm H}_2$ before it can lead to fragmentation \citep[e.g.,][]{2022MNRAS.510..177B}.

With observational constraints and theoretical seed models in tension, the only plausible alternative is to relax the assumptions on growth rate. Therefore, following \cite{2008MNRAS.386..989J}, we reformulate Equation \ref{eqn:simple_eddington} by replacing a generic $\epsilon$ with more specific values: $\epsilon_A$, $\epsilon_R$, and $\epsilon_J$. The first term ($\epsilon_A$) is the overall accretion efficiency and is simply the sum of the latter two---$\epsilon_A = \epsilon_R + \epsilon_J$---which correspond to the radiative accretion efficiency ($\epsilon_R$) and the jet accretion efficiency ($\epsilon_J$), respectively. Although all energy lost to either jets or radiation is mass-energy that does not contribute to the black hole's growth, jets are tightly collimated---in contrast to the isotropic radiation field of the accretion disk. As such, jets only take one hit out of the accretion rate, and we therefore can write a new growth equation,
\begin{equation} \label{eqn:complex_eddington}
\ln\left[({\rm M}(t)/{\rm M}(t_0)\right] \propto (t - t_0) \times (1 - \epsilon_A)/\epsilon_R.
\end{equation}

The power of Equation \ref{eqn:complex_eddington} is demonstrated in Figure \ref{fig:mass_growth}. 500 seed black holes were randomly initialized with masses between $10^2$ and $10^3\ {\rm M}_\odot$ at times of 75 to 175 Myr after the Big Bang, then allowed to evolve to redshifts of $z{\sim}7$. Every 10 Myr, these simulated AGN were randomly assigned to either be ``typical'' Eddington growers ($\epsilon_R=0.1$, $\epsilon_J=0.0$, 90\% probability each time step) or jet-assisted growers ($\epsilon_R=0.03$, $\epsilon_J=0.2$, 10\% probability). In contrast to the limits imposed by only assuming radiative effects in SMBH growth models, these light seeds are still capable of producing the most extreme quasars seen thus far \citep[see also][]{2008MNRAS.386..989J, 2013MNRAS.432.2818G}.

Although the methodology used to simulate growth tracks for Figure \ref{fig:mass_growth} is an ansatz model, it does point out the constraints needed to determine the feasibility of such growth. First, we need to understand what fraction of AGN at large redshifts are jet-dominated. Previous work has shown that the radio-loud fraction is consistent at ${\sim}10$\% up to $z \sim 6$ \citep{2015ApJ...804..118B}; however, as we show below, radio-loudness might not correlate with jetting at such extreme redshifts. Correspondingly, the values of $\epsilon_R$ and $\epsilon_J$---both for quasars in jet-dominated and radiation-dominated modes---are necessary components of these studies. Luckily, these can be derived from observations, being proportional to the AGN luminosity and jet power, respectively \citep{2008MNRAS.386..989J}.

\section{The Hidden Population}

Despite countless observations, the detection of extended jets (kpc-scale or larger) in the Epoch of Reionization through targeted radio campaigns has been limited in success. Very long baseline interferometric (VLBI) observations have identified compact, linear sources near quasars with total lengths of hundreds of parsecs \citep{2008A&A...484L..39F, 2020A&A...643L..12S,2008AJ....136..344M,2018ApJ...861...86M}. Longer sources---as are seen in abundance at low redshifts---are not seen in this era. The underlying problem is that at large scales the magnetic fields in jets are significantly smaller than close to the quasar itself, and so radio emission may not be how jets radiate in those locations.

The dearth of detected extended jets, particularly when considering the relative constancy of radio-loud quasars through this epoch, can be blamed on the changing conditions of the Universe in this era. The energy density of the Cosmic Microwave Background (CMB) scales as $\propto(1+z)^4$, which is, in turn, experienced by particles in the jet as being boosted by the Lorentz factor, $\Gamma$, squared \citep{2000ApJ...544L..23T,2001MNRAS.321L...1C}. At low redshifts, the CMB energy density is still less than that of magnetic fields, even in lobes. However, at $z>1$---and $z\gtrsim6$, in particular---this energy density will dominate over magnetic fields, potentially even in pc-scale jets! Jetted particles will therefore primarily radiate through inverse Compton interactions with CMB photons (IC/CMB) instead of radio synchrotron, rendering them X-ray dominant in their emission \citep[e.g.,][]{2014MNRAS.438.2694G}.

\begin{figure}[!t]
\centering
\includegraphics[]{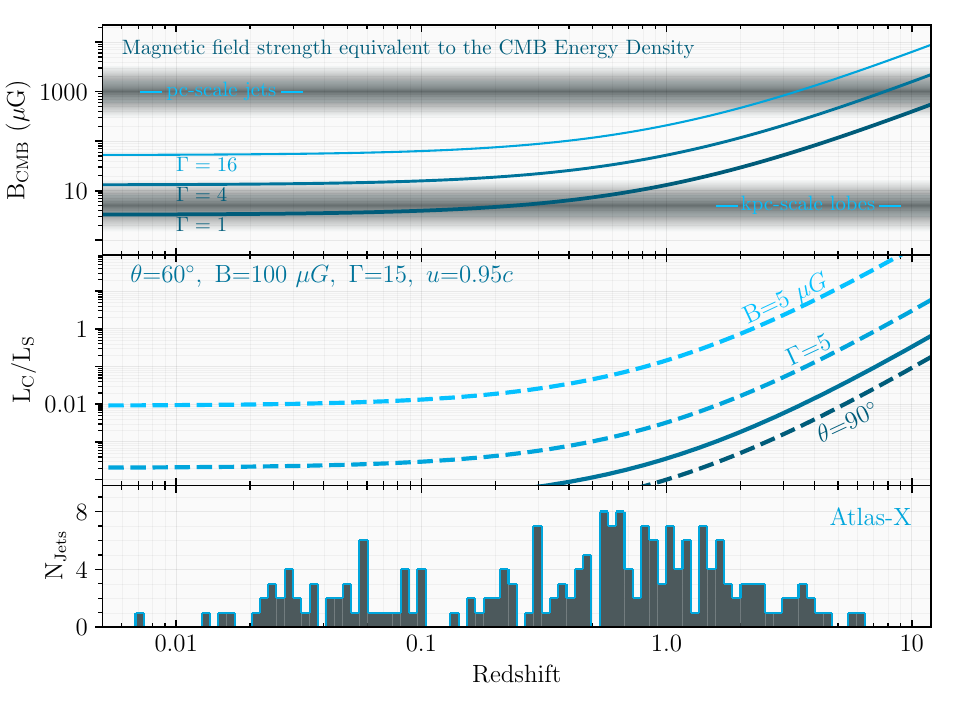}
\caption[Cosmic Evolution of CMB Contribution to Jet Emission]{{\bf Top}: The magnetic-field equivalent to the energy density of the CMB experienced by particles in a jet as a function of redshift. For increasing bulk Lorentz factor $\Gamma$ and probing into the Epoch of Reionization, the CMB becomes the dominant energy source for jetted particles to interact with. {\bf Center}: The relative luminosities of IC-CMB and Synchrotron for the IC-CMB model as a function of redshift. A fiducial model (parameters given in the top left) is shown with a solid line, while single-parameter variations are indicated with dashed lines. For strong magnetic fields, IC-CMB dominates at high redshift, while, for magnetic fields typical of lobes, the IC-CMB component dominates the radio synchrotron by orders of magnitude in that era. {\bf Bottom}: Histogram of the number of jets observed by \textit{Chandra}, as tracked by the Atlas-X compilation \citep{2023ApJS..265....8R}. Only a very limited number of jets have been observed to date even at $z>4$.}\label{fig:tripanel}
\end{figure}   

This redshift evolution is shown in Figure \ref{fig:tripanel}. The CMB energy density experienced by a particle with bulk Lorentz factor $\Gamma$ is comparable to a magnetic field with value ${\rm B}_{\rm CMB}=3.26 (1 + z)^2\Gamma\ \mu{\rm G}$ \citep{2017MNRAS.466.4299L}. As such, by $z \gtrsim 1$ the CMB will dominate over magnetic fields in large-scale structures. The overall balance between the synchrotron and IC/CMB luminosity is shown in the middle panel of Figure \ref{fig:tripanel}, with an ansatz set of parameters. Particularly for weaker magnetic fields and lower bulk Lorentz factors, IC/CMB {\it will} be the dominant emission mechanism in the Epoch of Reionization. Yet, as shown in the bottom panel, most of the jets observed by \textit{Chandra} are low-redshift \citep{2023ApJS..265....8R}; this is most likely due to both an observational bias to observe targets where jets have already been seen and an observatory bias where observations with the aging \textit{Chandra} are prohibitively expensive.

Indeed, only two large-scale AGN jets have been seen so far in the Early Universe. The first of these reported \citep{2021ApJ...911..120C} is associated with PJ352${-}$15 ($z=5.83$), the radio-loudest non-blazar quasar yet known in the Epoch of Reionization \citep{2018ApJ...861L..14B}. Previous VLBA observations had revealed a compact but extended radio source, with overall physical size of order 1 kpc \citep{2018ApJ...861...86M}, the largest such radio structure seen at these redshifts. Yet deep (265 ks) \textit{Chandra} observations revealed a second source, at the same position angle as the radio emission but approximately 50 kpc in projection from the quasar itself. With no corresponding radio or NIR emission, this was the first candidate for IC/CMB in the Epoch of Reionization ever found---but it required significant investments with \textit{Chandra}, and would have been impossible with \textit{XMM}'s angular resolution.

The second high-redshift quasar with observed X-ray jets is around a blazar at $z=6.1$ \citep{2022A&A...659A..93I}. Owing to the tight viewing angle, this emission is contiguous with emission from the core---to the resolution of \textit{Chandra}---but it extends in the same direction as VLBA-detected jets \citep{2020A&A...643L..12S} and, as with PJ352${-}$15, it traces jets significantly longer than what is seen in radio frequencies. Direct extrapolation from this source is difficult, as the X-ray emission from blazars is expected to differ from ``typical'' quasars, owing to viewing angle effects. However, the presence of a single blazar viewed at angle $\theta_v < \Gamma^{-1}$ implies the presence of $2\Gamma^2$ sources of similar intrinsic properties (including black hole mass and redshift) that are not seen at such favorable angles \citep[e.g.,][]{2013MNRAS.428.1449G}. The detection of further blazars at high redshifts could therefore reveal the presence of a significant population of further, jetted quasars!

Similar results also come from lower redshifts. A jet stretching half an arcminute in length was serendipitously detected near a $z=2.5$ quasar, but there was no corresponding radio counterpart at these scales \citep{2016ApJ...816L..15S}. The serendipitous nature of this discovery implies that many more such jets may be hiding, unseen because radio surveys did not see signs of a jet. Further, targeted searches have also found X-ray jets without radio emission around other $z > 2$ quasars \citep{2020ApJ...904...57S}, although there are still some radio knots coincident with X-ray emission at these lower redshifts \citep[e.g.,][]{2022ApJ...934..107S}. Nevertheless, in light of the dearth of detected large-scale radio emission in the Epoch of Reionization and the known physics of IC/CMB interactions, it is becoming clear that high-resolution X-ray imaging will be needed to make further progress on understanding the ubiquity and power carried by the earliest AGN jets.

\section{The need for AXIS}
Detecting and characterizing these jets requires X-ray observations. Even with the ngVLA or the SKA, the fundamental physical limitations imposed by the IC/CMB model will make large-scale jets unstudiable at radio frequencies. Likewise, \textit{JWST} or ALMA may be able to trace the inner-most regions of outflows through emission-line imaging of [O \RomanNumeralCaps{3}] or [C \RomanNumeralCaps{2}], respectively, but these are indirect tracers that would only indicate the presence of feedback, not its extent or power (to say nothing of their contamination from star formation). 

X-ray observations are thus clearly needed to study the earliest AGN jets, and the observatory capable of performing these investigations has two primary requirements. First, excellent angular resolution is necessary. Angular size scales at these redshifts are relatively flat at around 5 kpc/$^{\prime\prime}$, and so being able to separate jet emission from quasar emission beyond 10 kpc requires angular resolution of no worse than 2 arcseconds. Second, lobes may be extended \citep[e.g.,][]{2021AJ....161..207M} and will certainly be relatively faint, and so backgrounds must be as small as possible to enable statistically-robust detection. Beyond these requirements, higher effective area---particularly at soft X-ray energies---are needed.

\vspace{6pt} 

\acknowledgments{We kindly acknowledge the AXIS team for their outstanding scientific and technical work over the past year. This work is the result of several months of discussion in the AXIS-AGN SWG.}

\abbreviations{The following abbreviations are used in this manuscript:\\

\noindent 
\begin{tabular}{@{}ll}
AGN & Active Galactic Nucleus or Active Galactic Nuclei\\
AXIS & Advanced X-ray Imaging Satellite\\
CMB & Cosmic Microwave Background\\
IC-CMB & Inverse Compton emission from CMB photons\\
ngVLA & Next Generation Very Large Array \\
SKA & Square Kilometer Array \\
SMBH & Supermassive Black Hole \\
SWG & Science Working Group \\
VLBI & Very Long Baseline Interferometry
\end{tabular}}

\externalbibliography{yes}
\bibliography{references}

\end{document}